\documentclass[twocolumn,showpacs,preprintnumbers,amsmath,amssymb,PRB]{revtex4}
\usepackage{graphicx}
\usepackage{dcolumn}
\usepackage{bm}

\begin{document}

\title{From insulator to quantum Hall liquid at low magnetic fields}

\author{Tsai-Yu~Huang$^1$, C.-T.~Liang$^{1,\ast}$, Gil-Ho~Kim$^2$, 
C.~F.~Huang$^3$, Chao-Ping~Huang$^1$, Jyun-Ying~Lin$^1$, Hsi-Sheng~Goan$^{1,4}$, 
and D.~A.~Ritchie$^5$}
 \affiliation{$^1$Department of Physics, National Taiwan University, 
Taipei 106, Taiwan}
 \affiliation{$^2$School of Information and Communication Engineering and SAINT, Sungkyunkwan University, Suwon 440-746, Korea}
 \affiliation{$^3$National Measurement Laboratory, Center for Measurement Standards, Industrial Technology Research Institute, Hsinchu~300, Taiwan}
 \affiliation{$^4$Center for Theoretical Sciences, National Taiwan
University, Taipei 10617, Taiwan}
 \affiliation{$^5$Cavendish Laboratory, Madingley Road, Cambridge CB3 OHE, United Kingdom}

\date{\today}

\begin{abstract}
We have performed low-temperature transport measurements on a GaAs 
two-dimensional electron system at low magnetic fields. Multiple 
temperature-independent points and accompanying oscillations are observed in the 
longitudinal resistivity between the low-field insulator and the 
quantum Hall (QH) liquid. Our results support the existence of an intermediate regime, where
the amplitudes of magneto-oscillations can be well described by conventional Shubnikov-de Haas theory, 
between the low-field insulator and QH liquid.
\end{abstract}

\pacs{73.43.Qt, 73.40.-c, 73.21.La \\ 
$^{\ast}$Electronic mail: ctliang@phys.ntu.edu.tw
}
3
\maketitle

Magnetic-field-induced transitions in two-dimensional (2D) systems 
have attracted much interest 
recently \cite{klz,jiang,song,kim2}. To date, despite many existing experimental 
and theoretical studies on these transitions, there are still some 
interesting but unresolved issues. In particular, it is still under debate 
whether a direct transition from an insulator (I) to a high Landau 
level filling factor ($\nu\geq3$) quantum Hall (QH) state at low magnetic 
fields $B$ is a genuine phase transition. Experimentally, a 
\textit{single} approximately temperature ($T$)-independent point is observed in the 
longitudinal resistivity $\rho_{xx}$ at a transition magnetic field $B_{c}$ \cite{song}. 
For $B < B_{c}$, $\rho_{xx}$ decreases as the 
temperature is increased, reminiscent of an insulator. For $B>B_{c}$, 
$\rho_{xx}$ increases with increasing $T$, characteristic of a QH liquid. The 
low-field I-QH transition \cite{song} can be described by 
the destruction of the extended state within the tight-binding 
model \cite{song,weng}. However, Huckestein argued that such a low-field 
transition is not a phase transition, but can be identified as a crossover 
from weak localization to strong localization due to Landau quantization \cite{huckestein}. 
In his argument, the onset of QH liquid occurs as Landau quantization dominates with increasing 
$B$ and the crossover covers a very narrow region in $B$ so that it looks like a point. 
Nevertheless, recent experimental results \cite{tyh} demonstrate that the 
crossover from low-field insulator to Landau quantization can
cover a wide range of $B$ ($\sim$0.45~T), which indicates that we shall 
consider corrections to the argument raised by Huckestein \cite{cho,arapov}.
In fact, it has been shown that a QH liquid can be irrelevant to Landau quantization \cite{murzin1}, and
more studies are necessary to develop the related 
microscopic picture \cite{pruisken,murzin2}.

In addition to insulating and QH transport, actually the existence of a metallic regime \cite{klz,abrahams,nita,nita1,dubi1,dubi2} has been discussed
in the literature although such a regime may be unstable with respect to the disorder \cite{klz}. 
Based on the tight-binding model, there can be a metallic phase 
between the low-field insulator and QH liquid \cite{nita}. It is believed that electron-electron (e-e) interaction plays an 
important role in the appearance of a metallic phase \cite{abrahams}.  
Recently a unifying model for various 2D transport regimes is proposed 
by Dubi, Meir and Avishai \cite{dubi1,dubi2}. In their work, percolation \cite{trugman,shimshoni} and dephasing
are both considered. 
In this paper, we present an experimental to study various 2D transport regimes at low magnetic fields. 
In our case, multiple $T$-independent points and oscillations in $\rho_{xx}$ are observed 
between the low-field insulator and the QH liquid. By analyzing the amplitudes of 
these oscillations, we show the importance of metallic behavior described by conventional Shubnikov-de Haas 
(SdH) theory, which is derived by considering Landau quantization without resulting in strong localization \cite{coleridge}.
By changing the applied gate voltage, we are able to observe a single 
$T$-independent point in $\rho_{xx}$ which corresponds to the conventional 
low-field I-QH transition as reported previously \cite{song}. 
Our experimental results suggest that the insulating behavior and metallic
regime can coexist. Moreover, Landau quantization effect can modulate the density 
of states of a 2D system without causing the formation of QH liquid.

Our sample was grown by molecular-beam epitaxy and consists of a 20-nm-wide AlGaAs/GaAs/AlGaAs quantum 
well. The following layer sequence was grown on a GaAs (100) substrate: 50 nm undoped Al$_{0.33}$Ga$_{0.67}$As, 20 nm GaAs, 
40 nm undoped Al$_{0.33}$Ga$_{0.67}$As, 40 nm doped Al$_{0.33}$Ga$_{0.67}$As, and finally 17 nm GaAs cap layer. 
The growth of the 20-nm-wide GaAs quantum well was interrupted at its center, the wafer
was  cooled from 580~$^{0}$C to 525~$^{0}$C. The shutter over the 
In cell was opened for 80 sec, allowing the growth of 2.15 mono layer of InAs. 5 nm GaAs cap layer was then grown at 
530~$^{0}$C, before the substrate temperature was increased to 580~$^{0}$C for the remainder of the growth.
In our sample, the self-assembled InAs quantum dots, typically 4~nm in height and 28~nm in diameter, are formed near the center of the GaAs quantum well. Earlier studies on InAs quantum dots grown in the 
vicinity (15-80 nm) of a two-dimensional electron system (2DES) show that a dot-induced potential modulation 
drastically reduces the mobility \cite{sakaki}. When InAs quantum dots are placed close to the 2DES ($\leq 20$~nm), 
charging effects were observed \cite{horiguchi,yusa}. Studies on scattering introduced by InAs quantum dots grown 3~nm
from the GaAs/Al$_{0.33}$Ga$_{0.67}$As interface show that self-assembled InAs quantum dots can tailor the elctronic properties of a 2DES \cite{ribeiro}. It has been proved that a 2DES containing InAs quantum dots is suitable for studying low-field I-QH transitions \cite{kim2}.
The device was made into Hall pattern by standard 
lithography and etching processes and a NiCr/Au gate was evaporated on the surface.
At $V_g =0$~V, the carrier density of our 2DES is 1.4 $\times 10^{15}$~m$^{-2}$ with a mobility of $\approx 1.0$~m$^2$/Vs. Four-terminal magnetoresistivities were measured in a He$^{3}$ cryostat using standard ac phase-sensitive lock-in techniques with a current of 10 nA.

Figure 1 shows the longitudinal resistivity $\rho_{xx}\left(B\right)$ 
for $V_{g}=-0.02$~V at various $T$. At low $B$, 
the 2DES behaves as an insulator in the sense that $\rho_{xx}$ 
decreases as $T$ is increased.  The inset to Fig.~1 shows
$\rho_{xx}\left(B\right)$ and Hall resistivity $\rho_{xy}\left(B\right)$ for $V_{g}=-0.02$~V. 
At $B> 2.5$~T, we can see a well-developed $\nu=2$ QH state characterized by
the Hall plateau and zero longitudinal resistivity. Therefore the 
2DES behaves as an insulator at low fields and becomes a QH 
liquid with increasing $B$. Interestingly, between the QH state 
and the low-field insulator we observe four approximately 
$T$-independent points in Fig. 1 as indicated by arrows, in sharp contrast to only a 
single point reported in most experimental results.

The multiple $T$-independent points in $\rho_{xx}$ suggest an intermediate regime, which is neither weakly insulating nor quantum Hall-like, between the low-field insulator and QH liquid. The existence of such a regime is a key ingredient in the unifying model for various 2D transport regimes as reported in Ref.\cite{dubi1,dubi2}. We note that a 2DES in such a regime behaves as a metal, to which the e-e interaction effect characterized by the $T$-dependent Hall slope may be important\cite{abrahams}. As shown in the inset to Fig. 1, the Hall slope at $T=0.25$~K is higher than that at $T=1.6$~K, suggesting that the e-e interaction effect is present in our system.

It is well known that low-field Landau quantization in the metallic regime is well described by SdH formula\cite{coleridge}. 
In our case, the observed magneto-oscillations in the intermediate regime shown in figure 1 are periodic in 1/$B$, characteristics of Landau quantization effects. In order to test whether the conventional SdH theory, originally derived in the metallic region, is valid in the intermediate regime, we fit the amplitudes of the observed oscillations shown in Fig. 1 to the SdH formula \cite{coleridge,martin,brana}
\begin{equation}
   \bigtriangleup\rho_{SdH}(B,T) = C{\rm exp}(-\pi /\mu B)D(B,T),    
\end{equation}
with D(B,T)=(2$\pi^{2}k_Bm^{*}$T/$\hbar$eB)/sinh(2$\pi^{2}k_{B}m^{*}$T/$\hbar$eB). Here $\mu , m^{*}, T, k_{B}$, and $\hbar$ are the mobility, 
effective mass, temperature, the Boltzmann constant, and Plank 
constant divided by 2$\pi$, and $C$ is a constant independent of $B$ and $T$.
It is expected that $C$ is four times of the zero-field resistivity although there are reports on its deviations. As shown in the solid line in Fig. 2, the amplitudes $\Delta\rho_{SdH}$ 
are well described by Eq. 1. At the lowest temperature $T=0.25$~K, as shown 
in the inset to Fig.~2, 
ln$\bigtriangleup\rho_{SdH}$ is linear with respect to $1/B$, which is 
consistent with Eq.~(1) under which $\bigtriangleup\rho_{SdH} \propto 
\mathrm{exp}(-\pi /\mu B)$ when $T$ is so small that $D(B,T) \rightarrow 1$. 
Taking $\bigtriangleup\rho^{(0)}_{SdH}(B)$ as $\bigtriangleup\rho_{SdH}$ at 
$T=0.25$~K, we find that the ratio $\bigtriangleup\rho_ 
{xx}(B,T)/\bigtriangleup\rho^{(0)}_{SdH}(B)$ with $T>0.25$~K tends to decrease 
as $B$ increases, which is also consistent with Eq.~(1) of which the 
$T-$dependent factor $D(B,T)$ is a function of $T/B$. Moreover, using 
Eq.~(1), we find that at $V_{g}=-0.02$~V the measured effective mass 
$m^{*}={(0.0698 \pm 0.0012)}m_{0}$, close to the 
expected value $0.067m_{0}$ in GaAs. Here $m_{0}$ is the rest mass of 
a free electron. Our experimental results therefore clearly demonstrate that metallic behavior (magneto-oscillations governed by SdH theory) exists in this intermediate regime. 

According to the SdH theory, $\rho_{xx} \sim 
\rho_{0} + \Delta \rho _{SdH} \mathrm{cos} [\pi (\nu-1)]$, where $\rho_{0}$ represents the nonoscillatory part. 
In the original derivation, $\rho_{0}$ is taken as a constant and equal 
to the longitudinal resistivity at zero magnetic field. Thus a series 
of $T-$independent points are expected when cos$[\pi (\nu-1)]=0$, which 
occurs when 
\begin{equation}
\nu = n \pm \frac{1}{2},            
\end{equation}
where $n$ is an integer. The positions in $B$ of the four $T$-independent points labeled by arrows can be well described by Eq. (2), indicating that these points could be crossing points in the SdH oscillations. However, we would like to point out that this is only a possible explanation since the non-oscillatory background cannot be taken as a constant in our case. As shown in Fig.~3, with further increasing the gate voltage to $0$~V, we observed two more 
crossing points governed by Eq. (2). 
On the other hand, only three crossing points are observed between the 
low-field insulator and QH liquid as we decrease $V_g$ to -0.04~V as shown in Fig.~4.
While the positions in $B$ of $T$-independent points labeled by arrows are well described by Eq. (2),  
the crossing point indicated by a vertical line at the highest $B$  
is so close to a SdH maximum that its location does not follow Eq. (2). By further 
decreasing the gate voltage, as shown in Fig. 5, only such a crossing point is between the 
low-field insulator and QH liquid at $V_{g} =-0.06$~V. The inset (a) to Fig.~5 shows a zoom-in of 
$\rho_{xx}(B)$ near the crossing point at various $T$. We can see that the 
transition point is $T-$independent within the experimental errors.
Hence we can unify the prediction of the intermediate regime and the conventional I-QH transition. The 2DES enters a $\nu=4$ QH state directly from the insulator at the single transition point $B_{c}$, and such a transition is a direct 0-4 transition. Here the numbers 0 and 4 correspond to the insulator and $\nu=4$ QH state.   
In our study, $\rho_{xx}$ does not equal to $\rho_{xy}$ at $B=B_{c}$. However, we find that
$\mu B_{c} \sim 1$ where mobility $\mu$ is obtained from the observed SdH oscillations. Therefore, well-separated Landau bands are important to a direct transition in our study\cite{song} while Landau quantization can result in SdH oscillations
when $\mu B<1$\cite{martin}. We note that different mobilities should be introduced to understand the transport and 
universality\cite{song} near the low-field I-QH transition\cite{cho,arapov}.

As mentioned earlier, conventional SdH theory is derived based on Landau quantization
in the metallic regime. As shown in Fig. 5, for $B < B_{c}$ oscillations
in $\rho_{xx}$ are observed in the insulator when $V _{g} = -0.06$~V. The amplitudes of these oscillations are also well described by 
the SdH theory, as shown in open symbols and a dashed line in Fig. 2. 
We also find that the determined effective mass is close to the well-established value 0.067$m_0$. Our results indicate that the magneto-oscillations may follow the SdH formula in the low-field insulator, where there could be metallic puddles responsible for such a formula, though more studies are required for understanding the deviations of $\rho_{0}$ from the
zero-field resistivity\cite{hang,chen}. Hence metallic and insulator behavior
can coexist in our system.

As shown in the inset (b) to Fig.~5, at a sufficiently large negative gate voltage $V_{g} = -0.117$~V, the 2DES behaves as an insulator even at $\nu \sim 2$ and hence the corresponding QH state is destroyed by strong disorder. The destruction
of such a QH state, which is the most robust and lowest one under the unresolved spin-splitting in our study,
indicates that the 2DES is an insulator at all $B$. In such a case, we can still observe oscillations in $\rho_{xx}$ due to Landau quantization as reported before\cite{kim2,tyh}. 
Therefore, our study provides another piece of evidence to support that Landau quantization can modulate the density of
states of a 2D system without resulting in the formation of a QH liquid \cite{murzin1}. 
We note that the tight binding model and percolation theory may explain the coexistence of Landau quantization and insulator\cite{nita}.

In our gated 2DES, when $V_{g}$ is varied from 0 to -0.117~V, the carrier
density is changed from 1.4 $\times 10^{15}$~m$^{-2}$ to 
7.1 $\times 10^{14}$~m$^{-2}$, and the zero-field resistivity is 
changed from 2.26 k$\Omega$ to 57.7 k$\Omega$ at $T=0.25$~K. In our system, a small change
in gate voltage ($\approx 0.12$~V) can result in such a large change in
both $n$ and $\rho_{xx}$, therby causing a dramatic change in the observed
SdH pattern. These results may be related to the unique structure of our sample.  

In conclusion, we have observed multiple crossing points in $\rho_{xx}$ between the low-field insulator and 
QH liquid. Magneto-oscillations following conventional SdH theory can be observed before the appearance of a QH liquid
with increasing $B$. By decreasing the gate voltage, we 
observe the conventional transition so that there is only one transition 
point between the low-field insulator and QH liquid. Therefore we are able to obtain a unified picture which relates the prediction on the metallic regime to the direct I-QH transitions.

This work was funded by the NSC, Taiwan and the EPSRC, United Kingdom. 
G. H. Kim was supported by Samsung Research Fund, Sungkyunkwan University, 2007. 
H.S.G. acknowledges support from the focus group program of NCTS, Taiwan.

\centerline{Figure captions}

Fig. 1 Longitudinal resistivity $\rho_{xx}$ as a function of magnetic 
field $B$ for $V_{g}=-0.02$ V at various temperatures $T$. Four 
approximately $T$-independent points are indicated by arrows. The inset shows 
$\rho_{xx}(B)$ ($T=0.25$ K) and Hall resistivity 
$\rho_{xy}(B)$ at $T=0.25$ K (black solid line)
and $T=1.6$ K (red solid line) for $V_{g}=-0.02$ V.

Fig. 2 $\Delta \rho_{SdH}/D (B,T)$ as a function of $1/B$ for $V_{g}=-0.02$~V and $V_{g}=-0.06$~V. The full symbols denote the SdH amplitudes at $V_{g}=-0.02$ V when the temperature $T$=0.25, 0.52, 0.71, 0.90, 1.07, 1.30, and 1.60 K. The open symbols denote the amplitudes at $V _{g} =-0.06$ V when $T$=0.25, 0.32, 0.47, 0.61, 0.78, 0.93, 1.01, 1.22 and 1.37 K. The solid and dashed lines correspond to fits to $\bigtriangleup\rho_{SdH}/D(B,T)=C$exp$(-\pi /\mu B)$ with $C$ is a constant independent of $B$ and $T$ when $V _{g}$=-0.02 V and -0.06 V, respectively. The inset shows ln$\bigtriangleup\rho_{SdH}$ as a function of $1/B$ at $T=0.25$ K when $V _{g}$=-0.02 V. The solid line corresponds to
a fit to $\bigtriangleup\rho_{SdH} = C$exp$(-\pi /\mu B)$.

Fig. 3 $\rho_{xx}(B)$ for $V_{g}=0$ at various $T$. Six approximately $T$-independent points are indicated by arrows.

Fig. 4 $\rho_{xx}(B)$ for $V_{g}=-0.04$~V at various $T$.
Two crossing points are indicated by arrows and the crossing point at the highest $B$ is indicated by
a vertical solid line.

Fig. 5 Traces of $\rho_{xx}(B)$ for $V_{g}=-0.06$~V at different $T$.
The transition field $B_{c}$ is indicated by a vertical solid line. 
Inset (a) shows a zoom-in of $\rho_{xx}$ near the transition field $B_{c}$ at various temperatures.
From top to bottom at $B=0.9$~T, $T = 0.25, 0.32, 0.47, 0.61, 0.78, 0.93, 1.01, 1.22$,
and $1.37$~K. Inset (b) shows $\rho_{xx}(B)$ for $V_{g}=-0.117$~V at various $T$.

\end{document}